\documentstyle[emulateapj]{article}

\begin{document}

\submitted{To appear in the September issue of the Astronomical Journal}

\title{POSTSTARBURST MODELS OF LINERS}

\author{Yoshiaki Taniguchi, Yasuhiro Shioya, and Takashi Murayama}

\affil{Astronomical Institute, Graduate School of Science,
       Tohoku University, Aramaki, Aoba, Sendai 980-8578, Japan;
       tani@astr.tohoku.ac.jp, shioya@astr.tohoku.ac.jp,
       murayama@astr.tohoku.ac.jp}

\begin{abstract}
Since the discovery of low-ionization nuclear emission-line regions
in many galaxies (LINERs),
it has been recognized that they constitute a class of active 
galactic nuclei (AGNs) which are thought to be powered by gas 
accretion onto a central, supermassive black hole.
LINERs are observed in approximately one third of galaxies 
in the local universe and it has been often thought that 
they harbor an AGN-like central engine with moderate activity.
However, some LINERs show no direct evidence for AGNs such as
broad emission lines, radio jets, hard X-ray emission,
spectral energy distributions which
are inconsistent with starlight, and so on.
For such LINERs (a subset of type 2 LINERs), 
we present new poststarburst models which explain some of
their most important optical narrow emission-line ratios.
In these models, the ionization sources are 
planetary nebula nuclei (PNNs) with temperature of $\sim 10^5$ K
which appear in the late-phase evolution of intermediate-mass stars
with mass between $\approx 3 M_\odot$ and $\approx 6 M_\odot$.
Such PNNs left in a typical starburst nucleus can produce 
an H$\alpha$ luminosity of 
$L({\rm H}\alpha) \sim 10^{38}$ ergs s$^{-1}$ for typical poststarburst
LINERs and $\sim 10^{39}$ ergs s$^{-1}$ only in exceptionally bright cases.
The PNN phase lasts until the death of the lowest-mass stars formed 
in the starburst, which is  $\sim 5 \times 10^8$ yr for an assumed
lower limit of the initial mass function of 3 $M_\odot$. This long
duration appears consistent with the observed higher
frequency of occurrence of LINERs
if every galaxy could experience the
starburst activity several times in its life.
We therefore propose that some LINERs which show no direct evidence
for AGNs may be poststarburst nuclei powered by a cluster of PNNs.
\end{abstract}



\section {INTRODUCTION}

For the past two decades,
LINERs have been recognized as one of important classes of AGNs
(Heckman 1978; Stauffer 1982; Keel 1983a, 1983b
Filippenko \& Sargent 1985; V\'eron-Cetty \& V\'eron 1986;
Phillips et al.\ 1986a, 1986b; Maiolino \& Rieke 1995;
Ho, Filippenko, \& Sargent 1997a, 1997c). 
They are defined by the following criteria based on
optical emission-line intensity ratios (Heckman 1978);
1) $I$([O {\sc ii}]$\lambda$3727)/$I$([O {\sc iii}]$\lambda$5007)
$\geq 1$ where [O {\sc ii}]$\lambda$3727 is used to designate
the [O {\sc ii}]$\lambda\lambda$3726, 3729 doublet, and 2) 
$I$([O {\sc i}]$\lambda$6300)/$I$([O {\sc iii}]$\lambda$5007) $\geq 1/3$.
The most recent and comprehensive survey for activity in galactic 
nuclei conducted by Ho et al.\ (1997a)
has shown that LINERs make up a significant fraction 
($\approx$ 20 -- 30 \%) of all galaxies and hence they 
constitute the most populous class of AGNs; i.e.,
$\approx$ 50 -- 75 \% of AGNs are LINERs.
Therefore, it is very important to understand the origin of LINERs.

Possible origins of LINERs discussed previously are as follows:
a) Low-ionization analog of typical AGNs
(Ferland \& Netzer 1983; Halpern \& Steiner 1983;
Filippenko \& Halpern 1984;
Ho, Filippenko, \& Sargent 1993; 1997d),
b) powering by shocks driven either by superwinds or by
nuclear radio jets (Daltabuit \& Cox 1972;
Koski \& Osterbrock 1976; Fosbury et al.\ 1978; Heckman 1978; 
Dopita \& Sutherland 1995, 1996; Alonso-Herrero et al.\ 2000;
Sugai \& Malkan 2000),
c) photoionization by Wolf-Rayet stars (Terlevich \& Melnick 1985),
d) photoionization by hot O stars
(Filippenko \& Terlevich 1992; Shields 1992), and 
e) photoionization by old  post-AGB stars in elliptical galaxies
(Binette et al.\ 1994).
It is noted that the underlying Balmer absorption
due to many A-type stars weakens Balmer emission and thus
results in apparently higher [N {\sc ii}]$\lambda$6583/H$\alpha$
intensity ratios in some cases (e.g., Taniguchi et al. 1996);
note that this ratio is often used to
distinguish LINERs from H {\sc ii} regions; i.e.,
$I$([N {\sc ii}]$\lambda$6583)/$I$(H$\alpha$) $\geq 0.6$ for
LINERs while $< 0.6$ for H {\sc ii} nuclei
(Veilleux \& Osterbrock 1987; Ho et al.\ 1997a).
However, the overall statistics of the incidence of LINERs
(e.g., Ho et al. 1997a) are probably not affected by Balmer absorption
because careful starlight subtraction has been made to remove the
effects of Balmer absorption from the data,
by matching the stellar populations of each LINER with a
template absorption-line galaxy.

\subsection{Basic concept}
\begin{figure*}[b]
\epsscale{0.8}
\plotone{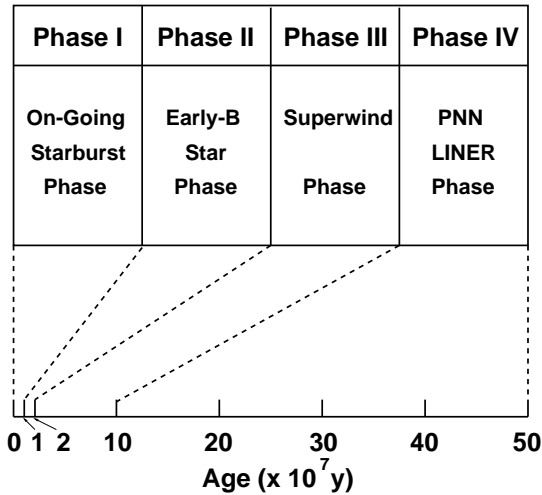}
\caption{%
A schematic illustration of the starburst evolution
proposed in this paper.
\label{fig1}}
\end{figure*}

There are several lines of evidence for the presence of pure
AGN in some LINERs:
i) the presence of broad emission-line region (BLR)
(Ho et al.\ 1997d; Maoz et al.\ 1998),
ii) the presence of hidden BLR in polarized optical spectra
(Wilkes et al.\ 1995; Barth, Filippenko,\& Moran 1999),
iii) a correlation between line width and critical density of
the forbidden lines arising from narrow emission-line region (NLR)
(Filippenko \& Halpern 1984),
iv) the presence of point-like ultraviolet source\footnote{
Since UV spectra have shown that some
of the point-like UV sources are in fact young starburst clusters
rather than AGNs (Maoz et al. 1998), the presence of an unresolved 
UV source may not always indicate the presence of an AGN.}
(Maoz et al.\ 1995),
v) the presence of radio core and/or jets (e.g., Heckman 1980;
Wrobel 1984; Slee et al. 1994),
and vi) the detection of hard X-ray continuum as well as  Fe K emission
(Iyomoto et al.\ 1996; Ishisaki et al.\ 1996;
Terashima et al.\ 1998a, 1998b, 2000;
Roberts, Warwick, \& Ohashi 1999).
Despite these firm lines of evidence for the presence of pure AGN
in a non-negligibly large number of LINERs (e.g., $\approx$
20\% of LINERs have the BLR: Ho et al. 1997d),
the majority of LINERs  have no firm evidence for the presence
of pure AGN, suggesting that LINERs are heterogeneous
(Heckman 1986).

In this paper, we investigate whether or not there are plausible
poststarburst models
for the ionization mechanism of LINERs
taking account that intermediate-mass stars ($\approx$ several
$M_\odot$) are also formed in nuclear starbursts (Joseph 1991).
Binette et al. (1994) showed that
old post-AGB stars provide sufficient ionizing photons to account
for the observed H$\alpha$ luminosity of LINERs associated
with elliptical galaxy nuclei. In their model, it was
assumed that post-AGB stars
were formed in the early phase of the galaxy evolution
(i.e., the initial starburst). On the other hand, in our paper,
our interest is addressed to starbursts occurred in recent pasts
in nuclei of spiral galaxies. Therefore, our model presented in this
paper is different from theirs.

\section{STARBURST AND POSTSTARBURST EVOLUTION}

In order to investigate the evolution of a star cluster
formed in a starburst, it is necessary to know both 
the initial mass function (IMF) and the star formation rate (SFR).
The SFR for nuclear starbursts can be estimated using
observed H$\alpha$ luminosities and so on (Kennicutt 1998).
Although it is generally difficult to derive the IMF accurately, 
it seems reasonable to adopt a power-law form of IMF;
$\phi(m) \propto m^{-\mu}$ (e.g., Scalo 1986).
In this formulation,
there are three free parameters; the power index ($\mu$), and 
the upper and lower mass limits of the IMF ($m_u$ and $m_l$).
The following values are often adopted for the evolution of
the solar neighborhood;
$\mu$ = 1.35, $m_l = 0.1 M_\odot$, and $m_u = 60 M_\odot$.
On the other hand, 
summarizing various kinds of observational constraints, 
Joseph (1991) suggests that $m_u \approx$ 30 -- 60 $M_\odot$ and 
$m_l \approx$ 3 -- 6 $M_\odot$ for typical nuclear starbursts.
Although the top-heavy IMF is suggested for the nuclear 
starbursts (i.e., $\mu$ is smaller than 1.35; e.g.,
Scalo 1990), we adopt $\mu$ =1.35,  $m_u = 60 M_\odot$, and 
$m_l = 3 M_\odot$ as a modest combination.
Another parameter is the duration of the starburst, $\tau_{\rm SB}$;
we adopt $\tau_{\rm SB} = 10^7$ yr because the negative feedback
from supernova explosions to star-forming gas clouds is expected
to occur $\sim 10^7$ yr (i.e., the lifetime of B stars)
after the onset of the starburst  (e.g., Larson 1987).
The lifetime of a starburst
can also be estimated as a gas consumption timescale,
$\tau_{\rm SB} \sim M_{\rm gas} \eta_{\rm SF} / SFR$ where $M_{\rm gas}$
is the gas mass available for the starburst, $\eta_{\rm SF}$ is the
star formation efficiency, and $SFR$ is the star formation rate.
If we adopt $M_{\rm gas} \sim 10^9 M_\odot$, $\eta_{\rm SF} \sim 0.1$,
and $SFR \sim 10 M_\odot$ yr$^{-1}$, we obtain $\tau_{\rm SB} \sim 10^7$ y.
We will discuss the SFR later.

\begin{figure*}[b]
\plotone{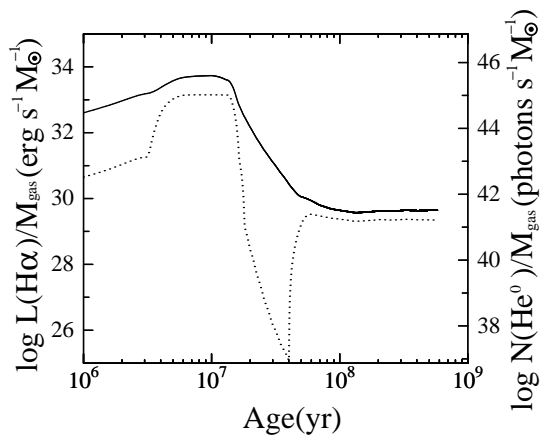}
\caption{%
The H$\alpha$ luminosity evolution of the star cluster
for the starburst models. Note that the H$\alpha$ luminosity is normalized
by a unit gas mass, $1 M_\odot$. The number of Lyman continuum photons
which are capable of ionizing He$^0$ (i.e., $h\nu \geq$ 24.6 eV)
is also plotted as a function of time by dotted lines. This is also
normalized by a unit gas mass, $1 M_\odot$.
\label{fig2}}
\end{figure*}

Now let us consider the evolution of the star cluster with the
above parameters. 
Phase I ($0 \leq t \leq 10^7$ yr): 
We assume that the starburst lasts for $10^7$ yr.
The star formation rate is assumed to be constant.
In this phase, main photoionization sources are
most massive stars. Dynamical effect of supernova
explosions may be weak in this phase. 
Phase II ($10^7 \leq t \lesssim 2 \times 10^7$ yr): 
Since the star formation ceases at a time
$t = 10^7$ yr in our model, main photoionization 
sources change from O stars to B stars as time goes. 
Wolf-Rayet stars also work in the photoionization
(Vacca \& Conti 1992; Conti 1999).
Phase III ($2 \times 10^7 \lesssim  t \lesssim 1 \times 10^8$ yr): 
Continuous supernova explosions develop a superwind and then
shock heating also works in this phase
(Heckman, Armus, \& Miley 1990; Ohyama, Taniguchi, \& Terlevich 1997;
Alonso-Herrero et al.\ 2000).
Phase IV [$t \simeq (1 {\em -} 5) \times 10^8$ yr]: 
Intermediate-mass stars
with mass of several $M_\odot$ play a role as 
the photoionization sources. 
The bolometric luminosity of each star in the main sequence
phase is $L_* \approx 1 \times 10^3 L_\odot$.
Each star evolves from the main
sequence to the asymptotic giant branch (AGB).
Following this AGB phase, each star will lose its gaseous envelope
and then a hot stellar core
appears and ionizes the surrounding gas,
making a so-called planetary nebula (Kaler 1985; Vassiliadis, \& Wood 1994). 
Since this core (i.e., a planetary-nebula nucleus; hereafter PNN)
is so hot  (i.e., $T_{\rm eff} \gtrsim 10^5$ K),
they become to be major photoionization sources.
Finally, stars with mass of $\approx 3 M_\odot$
die at $t \simeq 5 \times 10^8$ yr\footnote{Note that the ages of
intermediate-mass stars are estimated as $5.7 \times 10^8$ yr,
$2.5 \times 10^8$ yr, and $1.4 \times 10^8$ yr for
stellar masses of $3 M_\odot$, $4 M_\odot$, and $5 M_\odot$,
respectively using a relation of $\tau(m) = 1.2 \times 10^{10}
m^{-2.78}$ yr for $m < 10 M_\odot$ (Theis, Burkert, \& Hensler 1992).}
and then the effect of the nuclear starburst disappears
unless stars less massive than 3 $M_\odot$ were formed in the starburst.
The starburst evolution is summarized schematically in Figure 1.

In many previous studies, Phase IV has not been taken into account
seriously.  However, 
the temperature of PNNs is hot (i.e., $T_{\rm eff} \gtrsim 10^5$ K)
enough to ionize the surrounding nebula. More importantly,
this higher temperature can be responsible for the formation of
partly-ionized regions in which [O {\sc i}] emission is thought to
arise. If the ionization parameter
is as low as that for typical LINERs,
the optical spectrum of such a nebula is expected
to be quite similar to those of LINERs.
Note that each PNN has a bolometric luminosity
of $L_{\rm PNN} \approx  10^4 L_\odot$ (e.g., Vassiliadis, \& Wood 1994),
being comparable to that of a hot O star
(Filippenko, \& Terlevich 1992; Shields 1992).

\subsection{Poststarburst evolution model}
In order to demonstrate the importance of the PNN cluster,
we investigate the poststarburst evolution for the star cluster
discussed in section 2.1 using luminosity evolution models of Kodama \&
Arimoto (1997). In Figure 2, we show the time evolution of 
H$\alpha$ luminosity which is normalized by a unit gas mass, 
$1 M_\odot$. Here we assume that stars are formed from
the gas with the solar metallicity for simplicity.
The H$\alpha$ luminosity is estimated from 
the number of Lyman continuum photons;
$L({\rm H}\alpha) \simeq 1.36 \times 10^{-12} N({\rm Lyc})$
ergs s$^{-1}$ (Leitherer \& Heckman 1995).
Note that the lower mass cutoff is fixed at 0.1 $M_\odot$
in Kodama \& Arimoto's models. Therefore, the H$\alpha$ luminosity 
shown in Figure 2 is corrected for the case of $m_l = 3 M_\odot$ by us.

According to Kennicutt (1998),
the SFR is related to $L({\rm H}\alpha)$ as
$SFR = 7.9 \times 10^{-42} L({\rm H}\alpha) ~ M_\odot {\rm yr}^{-1}$
for the IMF with $\mu = 1.35$, $m_l = 0.1 M_\odot$, and 
$m_u = 100 M_\odot$. This relation can be replaced by
$SFR = 4.0 \times 10^{-42} L({\rm H}\alpha) ~ M_\odot {\rm yr}^{-1}$
for the Salpeter IMF with $m_l = 3 M_\odot$ and $m_u = 60 M_\odot$. 
Balzano (1983) found $L({\rm H}\alpha) \simeq 10^{40}$ --
$6 \times 10^{42}$ ergs s$^{-1}$ for her sample of Markarian
starburst nuclei with a Hubble constant of $H_0$ = 75 km s$^{-1}$
Mpc$^{-1}$ (see also Kennicutt, Keel, \& Blaha 1989;
Ho, Filippenko, \& Sargent 1997b).
Since the duration of H$\alpha$-bright phase is very short 
(e.g., $\lesssim
10^7$ yr; see Figure 2) in the starbursts, it seems rarer to
detect starbursts with such a bright phase from a statistical point
of view. For example, 
the age of the starburst occurring in Mrk 1259, which is 
one of the brightest starburst nuclei studied by Balzano (1983),
is estimated to be $\sim 5 \times 10^6$ yr (Ohyama et al. 1997).
Although the observed
H$\alpha$ luminosity of this galaxy is  $\sim 10^{41}$ ergs s$^{-1}$,
this galaxy would be observed as a less luminous starburst
if it will be observed $10^{7-8}$ yr after now. 
Typical H$\alpha$ luminosities
listed in Balzano's catalog are $10^{40}$ - $10^{41}$ ergs s$^{-1}$.
If the above statistical effect is taken into account, 
their initial H$\alpha$ luminosities may be of the order of
$10^{41}$ - $10^{42}$ ergs s$^{-1}$. 
Furthermore, since the mean reddening correction is about a factor of
4-5 for H$\alpha$ (e.g., Balzano 1983), 
the reddening-corrected, initial H$\alpha$ luminosities seem to be as luminous 
as $\sim 10^{42}$ ergs s$^{-1}$ on average and some brightest nuclear starbursts
may have H$\alpha$ luminosities of $\sim 10^{43}$ ergs s$^{-1}$.
These values lead to $SFR \sim$ 10 -- 100 $M_\odot$ yr$^{-1}$.
Then, we estimate $M_{\rm gas} \sim 10^{8 - 9} M_\odot$
for the nuclear starbursts.
Indeed, the molecular gas mass in circumnuclear regions of
typical starburst nuclei is $M_{\rm gas} \sim 10^9 M_\odot$
(Devereux et al.\  1994).
As shown in Figure 2, the H$\alpha$ luminosity density per unit mass
at $t \sim 10^8$ yr is $L({\rm H}\alpha)/M_{\rm gas} 
\sim 10^{30}$ ergs s$^{-1}$ $M_\odot^{-1}$. Therefore,
we estimate typical H$\alpha$ luminosities of poststarburst 
nuclei with $t \sim 10^8$ yr as $L({\rm H}\alpha) \sim
10^{38}$ ergs s$^{-1}$ for a typical LINER powered by PNNs, 
and $\sim 10^{39}$ ergs s$^{-1}$ in exceptionally bright cases.

It should be mentioned that our poststarburst models cannot be
applied to on-going starburst galaxies because most starburst
nuclei have $L({\rm H}\alpha) \gtrsim 10^{39}$ ergs s$^{-1}$
(Balzano 1983; Kennicutt et al. 1989; Ho et al. 1997b).
Namely, if a nuclear starburst occurs, the photoionization 
should be dominated by massive stars in the starburst
even if there is a cluster of PNNs left from the recent past
nuclear starburst.

\begin{figure*}[t]
\plotone{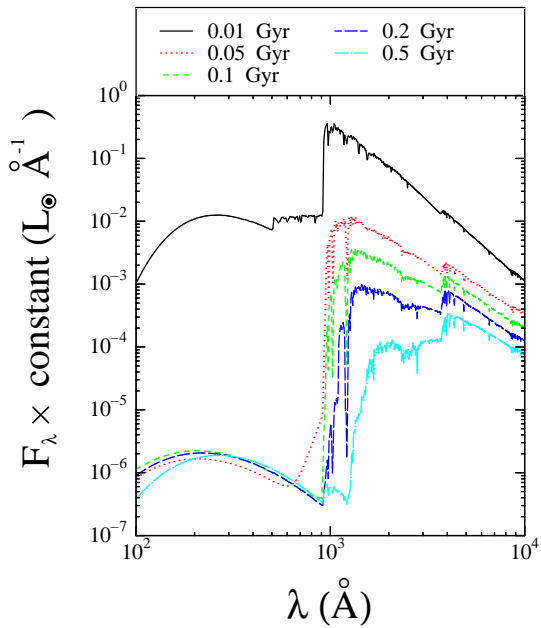}
\caption{%
The SED evolution of the star cluster adopted in our study
for $t$ = 0.01, 0.05, 0.1, 0.2, and 0.5 Gyr.
\label{fig3}}
\end{figure*}

\subsection{Optical emission-line properties of the ionized nebula}

Next we investigate optical emission-line properties of the gaseous
nebula photoionized by the PNN cluster.
We use the photoionization code CLOUDY version 90.05 (Ferland 1996),
which  solves the equations of statistical and
thermal equilibrium and produces a self-consistent
model of the run of temperature as a function of depth into the nebula.
Here we assume that a uniform-density gas
cloud with plane-parallel geometry is irradiated by 
the star cluster formed in the starburst.
The parameters for the calculations are 1) the hydrogen density
of the cloud ($n_{\rm H}$),
2) the ionization parameter (Osterbrock 1989);
$U = Q({\rm H^0})
(4 \pi r^2 n_{\rm H} c)^{-1}$ where $Q({\rm H^0})$ is the number
of ionizing photons, $r$ is the distance from the ionizing source,
and $c$ is the light velocity, 3) the spectral energy
distribution (SED) of the ionizing radiation,
and 4) the chemical compositions.

The SED of the ionizing radiation is calculated during the
course of the starburst evolution using Kodama \& Arimoto's
models. The SED evolution is shown in Figure 3.
Note that the ordinate is the flux in units of $L_\odot$ \AA$^{-1}$,
corresponding to the case that the total gas mass is $1 M_\odot$.
We perform our photoionization calculations
as a function of the ionization parameter between log $U = -4$ and
log $U = -3$ with a logarithmic interval of 0.2.
Following Filippenko \& Terlevich (1992), 
we adopt $n_{\rm H} = 10^3$ cm$^{-3}$.
In order to examine the effect of dust grains on 
emission-line intensity ratios, we perform photoionization 
calculations for the following two cases; a) the dust-free gas with
the solar abundances,
and b) the gas with the Orion Nebula abundances where effects of dust
grains are taken into account.
For both the cases we simply adopted the abundance sets available
in CLOUDY.
The solar composition is taken from
Grevesse \& Anders (1989) and Grevesse \& Noels (1993).
The Orion Nebula abundances
are a subjective mean of the abundances determined by 
Baldwin et al. (1991), Rubin et al. (1991), and
Osterbrock, Tran, \& Veilleux (1992).
The grains are the large-R grains described by Baldwin et al. (1991). 
The calculations were stopped when the temperature fell to 3000 K,
below which little optical emission is expected.

The results are shown in Figures 4 and 5 for the above two cases,
respectively. Here we use 
the same emission-line diagnostic diagrams as those used in 
Filippenko \& Terlevich (1992).
It is shown that our poststarburst models with age of 
$t \simeq (1$ -- $5)\times 10^8$ yr appear consistent with
the observed emission-line properties of LINERs; the depletion of 
metals into dust grains leads to higher relative intensities
of low-ionization lines such as [O {\sc i}] and [S {\sc ii}],
being more consistent with the observations.
However, it should be noted that the [O {\sc i}]/H$\alpha$ ratio 
barely touches the LINER region of the diagnostic diagram
in both Figure 4 and 5.  In this respect, our models are
similar to the previous O-star models for weak [O {\sc i}] LINERs
proposed by Filippenko \& Terlevich (1992) and Shields (1992);
i.e., the weak [O {\sc i}] LINERs are defined as LINERs with
$I$([O {\sc i}]$\lambda$6300)/$I$(H$\alpha$) $\leq 1/6$.
However, since PNNs are hot enough to supply such high-energy photons,
the other emission-line ratios appear more consistent 
with the observations (see Figures 4 and 5) than those of the O-star models.
The weakness of [O {\sc i}] emission in our models may be attributed
to the fact that the production rate of high-energy photons is lower than
that of the non-thermal continuum radiation from an AGN.

\begin{figure*}
\epsscale{1.2}
\plotone{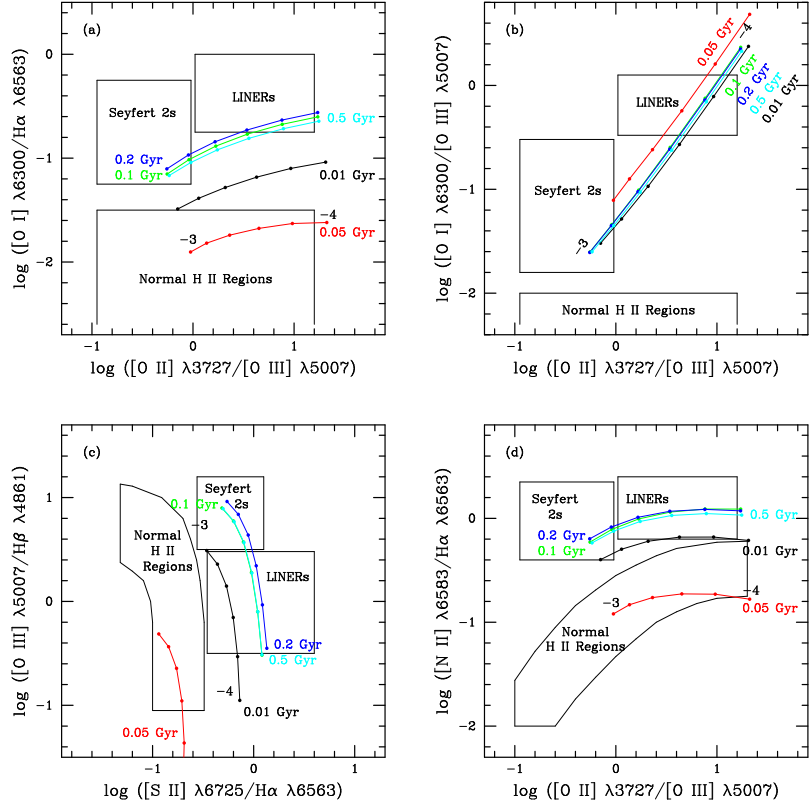}
\caption{%
Optical emission-line diagnostics used to classify different
types of emission-line galaxies adopted in Filippenko \& Terlevich (1992).
Typical loci of ratio combinations used in Filippenko \& Terlevich (1992)
are reproduced for normal H {\sc ii} regions, NLRs
of typical AGNs such as Seyfert 2 galaxies, and
classical LINERs. Results of poststarburst photoionization models for
$t$ = 0.01, 0.05, 0.1, 0.2, and 0.5 Gyr
are marked as line segments joining the points
between log $U = -4$ and log $U = -3$.
\label{fig4}}
\end{figure*}

\begin{figure*}
\epsscale{1.2}
\plotone{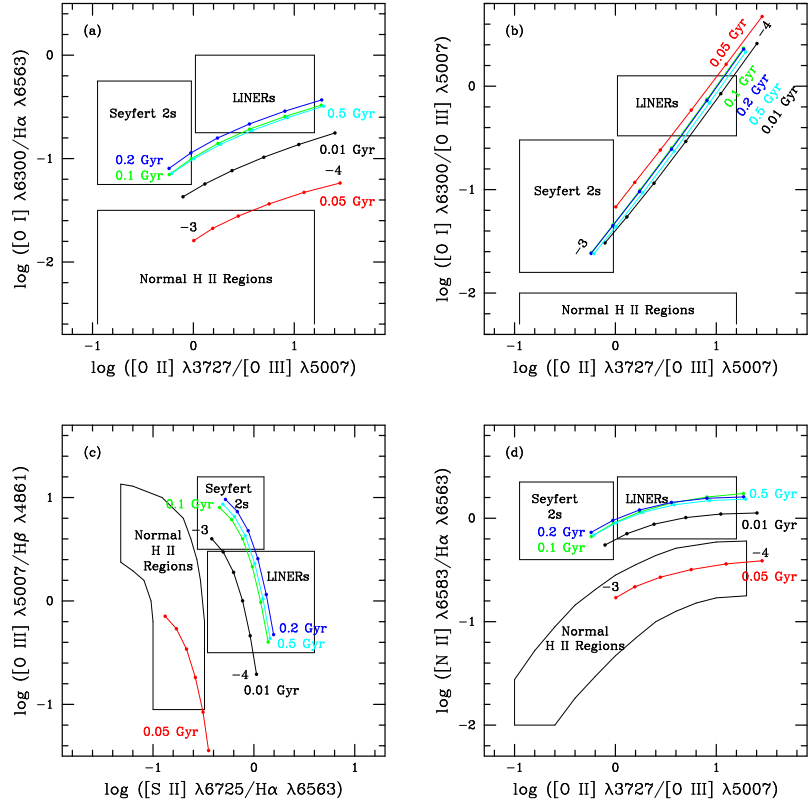}
\caption{%
The same diagnostic diagrams as those in Figure 4,
but for the Orion Nebula abundances.
\label{fig5}}
\end{figure*}

\begin{figure*}[b]
\epsscale{0.8}
\plotone{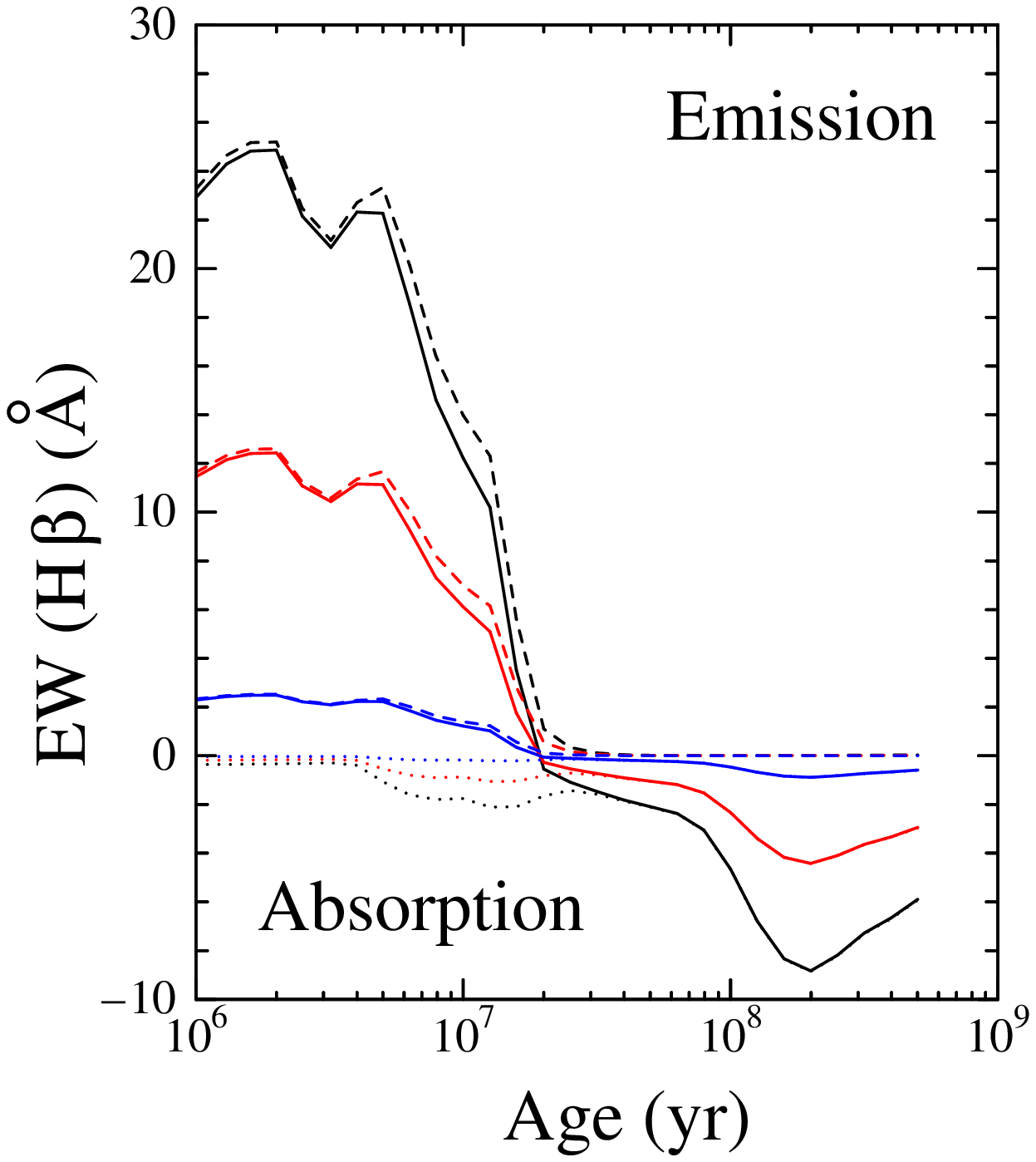}
\caption{%
The evolutions of equivalent width of both H$\beta$ emission and
absorption lines are shown for the three cases; 100 percent,
50 percent, or 10 percent of the blue continuum emission
arises from stars formed in the starburst.
Gaseous H$\beta$ emission, stellar H$\beta$ absorption, and their
sum total are shown by dashed lines, dotted lines, and thick lines,
respectively, for the three cases.
Here in order to estimate the equivalent width of H$\beta$ absorption,
we have used another spectral synthesis code GISSEL96
(Bruzual 1993; Leitherer et al.\ 1996)
because Kodama \& Arimoto's models do not give it. 
Since the two codes use both the same stellar evolution models
and stellar atmosphere models, there is no problem in the estimate.
The equivalent width is defined as 
$EW = \int {{F_{\rm line}(\lambda) - F_{\rm cont}(\lambda)}
\over F_{\rm cont}(\lambda)} d\lambda$ \AA. 
\label{fig6}}
\end{figure*}

In order to demonstrate this quantitatively, we show
the time evolution of the number of
ionizing photons which are capable of ionizing He$^0$ (i.e.,
$h\nu \geq$ 24.6 eV) in Figure 2 (see the dotted line).
Although high-energy photons are supplied by
O stars in Phase I, they decrease in number significantly
in Phase III. However, they increase in number again in Phase IV 
because of the supply from high-temperature PNNs. 
This makes our models more consistent with the observations
than the previous O-star photoionization models.

It is known that intermediate-mass stars produce strong
Balmer absorption lines in their stellar atmospheres and thus
the Balmer emission lines arising from the ionized nebulae 
are sometimes quenched by the Balmer absorption.
In order to investigate this effect quantitatively along the 
starburst evolution, we show time variations of the equivalent width
($EW$) of both Balmer emission and absorption for the case of H$\beta$
in Figure 6. The H$\beta$ emission dominates in the first 
a few times  $10^7$ yr while the H$\beta$ absorption does
after this age. This comparison suggests that LINERs 
whose Balmer lines are observed as absorption are more
popular by about one order of magnitude 
than those with the Balmer emission. On the other hand,
observations show that the fraction of the former LINERs 
is significantly smaller than that of the latter ones (Ho et al.\ 1997a).
However, if the Balmer emission has
a narrower line width on average than the Balmer absorption,
the Balmer emission can be seen even if the $EW_{\rm absorption} >
EW_{\rm emission}$. This trend is actually seen in the observed spectra
of many LINERs (Ho et al.\ 1997a).

\section{CONFRONTATION WITH OBSERVATION}

\subsection{Frequency of occurrence of poststarburst LINERs}

We estimate the frequency of occurrence of poststarburst
LINERs. Since LINERs are very common (Ho et al.\ 1997a),
any models are required to explain this property.
The lifetime of the poststarburst LINER would be as long as
$\sim 10^{10}$ yr if $m_l$ is less than 1 $M_\odot$ (Binette 
et al. 1994). However, this case is beyond our scope and should
be applied to the nuclear emission-line regions of
long-lived elliptical galaxies. Here we adopt that 
the lifetime of LINER in our poststarburst model is
$\tau({\rm LINER}) \simeq 5 \times 10^8$ yr given that 
the lowest mass of stars formed in the starburst is 3 $M_\odot$.
On the other hand, the lifetime of the starburst is 
$\tau^{\rm eff}({\rm SB}) \simeq 2 \times 10^7$ yr
where $\tau^{\rm eff}({\rm SB})$ is the sum total of the 
durations of both Phase I and II.
Therefore, the expected number ratio between starbursts and
poststarburst LINERs is $N({\rm SB})/N({\rm LINER}) \sim
\tau^{\rm eff}({\rm SB})/\tau({\rm LINER}) \sim 1/25$. 
Since typical nuclear-starburst galaxies are observed in 
$\approx$ 1\% of nearby galaxies (e.g., Balzano 1983),
poststarburst LINERs are expected to share $\approx$ 25\% of 
nearby galaxies, being consistent with the observation (Ho et al.\ 1997a).
However, we should mention why nuclear starbursts with
$\tau^{\rm eff}({\rm SB}) \sim 2 \times 10^7$ yr are observed 
in $\sim 1$\% of nearby galaxies because the comparison between
$\tau^{\rm eff}({\rm SB})$ and the Hubble time, $\sim 10^{10}$ yr, 
would give statistically a frequency of
occurrence of starbursts of $\sim$ 0.2\%.
The difference between this estimate and the actual frequency 
may be reconciled if every galaxy experiences starbursts
several times in its life.

We mention that the above estimate is based on an assumption that 
there has been no star formation activity since the recent past starburst.
If the galaxy currently experiences nuclear star formation at a level
of even $10^{-3}$ times the star formation rate of the starburst,
then the young stellar population would dominate over PNNs by an order
of magnitude, and the emission-line spectrum would be that of an 
H {\sc ii} region, not a LINER.
If there is any current or very recent star formation, 
the frequency of LINERs generated by this poststarburst mechanism 
may be significantly lower than the above estimate.
However, it has been argued that the negative feedback
from supernova explosions to star-forming gas clouds prevent from
new intense star formation for $\sim 10^8$ yr after the
onset of the starburst  (e.g., Larson 1987).
Therefore, our assumption seems reasonable. 

\subsection{Morphological-type segregation}

First of all, we mention that our models cannot be applied to LINERs
associated with the nuclei of elliptical galaxies and most of S0
in which nuclear starburst occurs seldom although some S0 galaxies
have a lot of molecular gas together with a moderate level
of star-formation activity (e.g., Thronson et al. 1989). 

It is known that there is a morphological-type segregation
between nuclear starbursts and LINERs (Ho et al.\ 1997a);
i.e., starbursts 
appear preferentially in late-type spirals while LINERs do in  
early-type ones. As for this issue,
we give the following three comments.
1) A possible activity classification bias: 
If one finds evidence both for an AGN and for a nuclear starburst
in a galaxy, one tends to classify it as an AGN. It is noted that
infrared evidence for nuclear starbursts has been found
even in a significantly large number of early-type spiral galaxies
Devereux 1987, 1994).
2) A possible spectroscopic bias: Emission-line classification based
on optical spectroscopy with a large aperture may be sometimes
unreliable. For example, even if a galaxy has a pure AGN,
it may be misclassified as a nuclear-starburst galaxy if it has very
luminous circumnuclear star-forming regions (Yoshida et al.\ 1993).
This means that one may misidentify activity types of 
late-type spirals with an AGN.
And, 3) a morphological-evolution effect: 
It seems likely that a strong nuclear starburst may modify
the appearance of a galaxy so that the host tends to be classified as
an earlier-type spiral (Alonso-Herrero et al.\ 2000;
Wada, Habe, \& Sofue 1995).
If starbursts were triggered by minor-mergers of satellite galaxies,
the morphology of host galaxies could be altered
(Mihos \& Hernquist 1994; Hernquist \& Mihos 1995; Taniguchi \& Wada 1996).

\section{SUMMARY}

We have presented a new poststarburst model of LINERs.
In this model, the ionization sources are planetary nebula
nuclei (PNNs). 
Our main point is that the ionization sources are
planetary nebula nuclei (PNNs) with temperature of $\sim 10^5$ K
which appear in the late-phase evolution of intermediate-mass stars
with mass between $\approx 3 M_\odot$ and $\approx 6 M_\odot$.
Our models are able to reproduce the observed optical narrow 
emission-line ratios of LINERs although the [O {\sc i}] emission 
is underpredicted to some extent.

We give a summary of the limitations of our models. 
1) Our models cannot be applied to LINERs associated with nuclei of 
elliptical galaxies and most of S0 galaxies in which nuclear starbursts
occur seldom. 
2) Our models cannot be applied to LINERs with direct evidence for
AGN such as broad-line emission, radio jets, hard X-ray emission,
and so on (mostly type 1 LINERs). 
3) Our models cannot be applied to on-going starburst nuclei because 
the photoionization is dominated by massive stars in the starburst
rather than a cluster of PNNs left from a recent past starburst.
4) Accordingly, our models are applied to a subset of type 2 LINERs 
located in nuclear regions of spiral galaxies. 
Since the H$\alpha$ luminosity in the poststarburst LINER phase is
less luminous by $\sim$ 4 orders of magnitude than the initial
H$\alpha$ luminosity at the onset of the starburst, our models 
are preferentially applied to low-luminosity LINERs; e.g., 
$L$(H$\alpha$) $\sim 10^{38}$ ergs s$^{-1}$.

Our models are constructed to explain the observed optical narrow
emission-line ratios and we have not examined properties at other
wavelengths (e.g., radio continuum, hard X rays, and so on). 
However, since the ionization
sources are PNNs, poststarburst LINERs will not show any evidence for
the presence of AGNs such as radio jets, hard X-ray emission and so on. 
As shown in Figure 3, our models predict that the X-ray emission 
is much weaker than that of typical AGNs, being consistent with
some type 2 LINERs (Terashima et al. 2000 and references therein;
see also for AGN-like SEDs of LINERs, Ho 1999). 
One interesting prediction of our models is that the ionization
sources may be spatially extended because massive stars in nuclear 
starbursts are often distributed within central 100-pc regions (Meurer
et al. 1995; see also Sugai \& Taniguchi 1992).
Recently, Pogge et al. (2000) have shown that 
some LINERs have spatially-extended emission-line regions
with sizes of tens to hundreds parsecs.
However, since they have either a compact UV source or 
no bright UV source, they may be not poststarburst LINERs
but genuine low-ionization AGNs. 
In conclusion, in order to examine how many
poststarburst LINERs are really present, we need systematic investigations 
of radio-continuum and X-ray properties of low-luminosity LINERs
as well as UV continuum imaging in future.

\vspace{0.5cm}

We would like to thank an anonymous referee for
many useful comments and suggestions.
YS and TM are supported by JSPS.
This work was financially supported in part by Grant-in-Aids for the Scientific
Research (Nos. 10044052, and 10304013) of the Japanese Ministry of
Education, Culture, Sports, and Science.


\end{document}